\documentclass{iau}
\usepackage{graphicx}

\title[Multiple bursts of star formation in young star clusters: The case of the Orion Nebula Cluster] %% give here short title %%
{Multiple bursts of star formation in young star clusters:\\ The case of the Orion Nebula Cluster}

\author[Tereza Jerabkova]   %% give here short author list %%
{Tereza Jerabkova$^{1,2,3}$
}

\affiliation{$^1$
European Southern Observatory,\\ 
Karl-Schwarzschild-Strasse 2, 85748 Garching bei M\"{u}nchen\\
e-mail: {\tt tjerabko@eso.org} \\[\affilskip]
$^2$
Helmholtz Institut f\"{u}r Strahlen und Kernphysik, Universit\"{a}t Bonn,\\
Nussallee 14–16, 53115 Bonn, Germany \\[\affilskip]

$^3$
Astronomical Institute, Charles University in Prague,\\
V Hole\v{s}ovi\v{c}k\'ach 2, CZ-180 00 Praha 8, Czech Republic \\[\affilskip]}

\pubyear{2019}
\volume{351}  
\jname{Star Clusters: From the Milky Way to the Early Universe}
\editors{A. Bragaglia, M.B. Davies, A. Sills \& E. Vesperini, eds.}

\begin{document}

\maketitle
%. CONTINUE EDITING FROM HERE

\begin{abstract}
Young star clusters (YSCs) with resolved stellar populations are well suited for studying   star-cluster  formation.  In  most  cases,  the  (pre-main-sequence) stellar populations found in the YSCs are coeval with an intrinsic age spread of up  to  1Myr.  Such  observations  can  be  understood  as  the  YSCs  having  formed  in  one  burst, which star formation was truncated by stellar feedback. The recent discovery that the colour-magnitude diagram of the Orion Nebula Clusters (ONC) contains three well defined age-separated populations appears to  shatter  this  model.  The implication is that the ONC formed in three bursts, with star formation still on-going in the last burst. We present new observational results focusing on the three populations in the ONC using OmegaCAM photometry and Gaia DR2 measurements. We also describe a theoretical model which may explain these observations by an interplay between stellar feedback and cluster dynamics.

\end{abstract}

\firstsection 
\section{Introduction}
Up until recently much of the theoretical and observational research focused on the 
study of still embedded young star clusters (YSCs) yielded evidence that
the stellar population residing in YSCs is mostly coeval (with an age spread $<$ 1 Myr,
e.g. \cite{Lada2003}). But there is an ongoing discussion about a possibly larger age spread in some YSCs of up to several Myr (e.g. \cite{Palla2000}).
It is still to be understood whether the observed age spreads are real, being related to the physics of YSC formation, 
or if they are due to an inaccurate evaluation of the impact of observational biases such as differential extinction, stellar variability and processes related to proto-stars such as episodic accretion (e.g. \cite{Jeffries2011}).

Nevertheless, these observations, in which YSCs form within a 1-few Myr long burst of star formation, can be understood with a feedback-regulated model: once a YSC forms massive stars, their winds and UV radiation destroy surrounding gas and dust and thus truncate star formation (\cite[Kroupa, Jerabkova, et al. 2018]{Kroupa2018} and references therein).

In contrast to this understanding of the formation of YSCs, \cite{Beccari2017} reported the detection of three well separated sequences of pre-main sequence (PMS) stars in an optical colour-magnitude diagram (CMD) in an area of 1.5deg radius centered on the ONC, an observation which completely shakes our understanding of YSC formation. The stars belonging to the three sequences, while being all centered on the Trapezium, show a different spatial distribution with the apparently oldest (and most numerous) population being more spatially spread around the center of the ONC with respect to the youngest one which shows a stronger concentration toward the center. The effect of unresolved binaries and differential extinction was carefully considered, so that the most likely explanation
of the observed sequences is that there are three stellar populations of about 3, 2 and 1 Myr age in the ONC.

In this contributions I summarize the follow-up study of this region using Gaia DR2 data \cite{Jerabkova2019} and suggest a possible theoretical explanation based on \cite{Kroupa2018} and \cite{Wang2019}.

\section{Gaia DR2 view of stellar populations around the ONC}
In \cite{Jerabkova2019} we use Gaia DR2 data in combination with deep OmegaCAM photometry to perform a more detailed study of the three populations in the ONC.
We took advantage of parallax and proper motion values from the Gaia catalog to separate the ONC stellar population from the back/fore-ground contamination. On the other hand, Gaia DR2 photometry 
is not reliable in crowded regions such as the ONC, thus we used photometry from OmegaCAM using the $r$ and $i$ filters. 

\begin{figure*}[ht]
\begin{center}
 \includegraphics[width=5.2in]{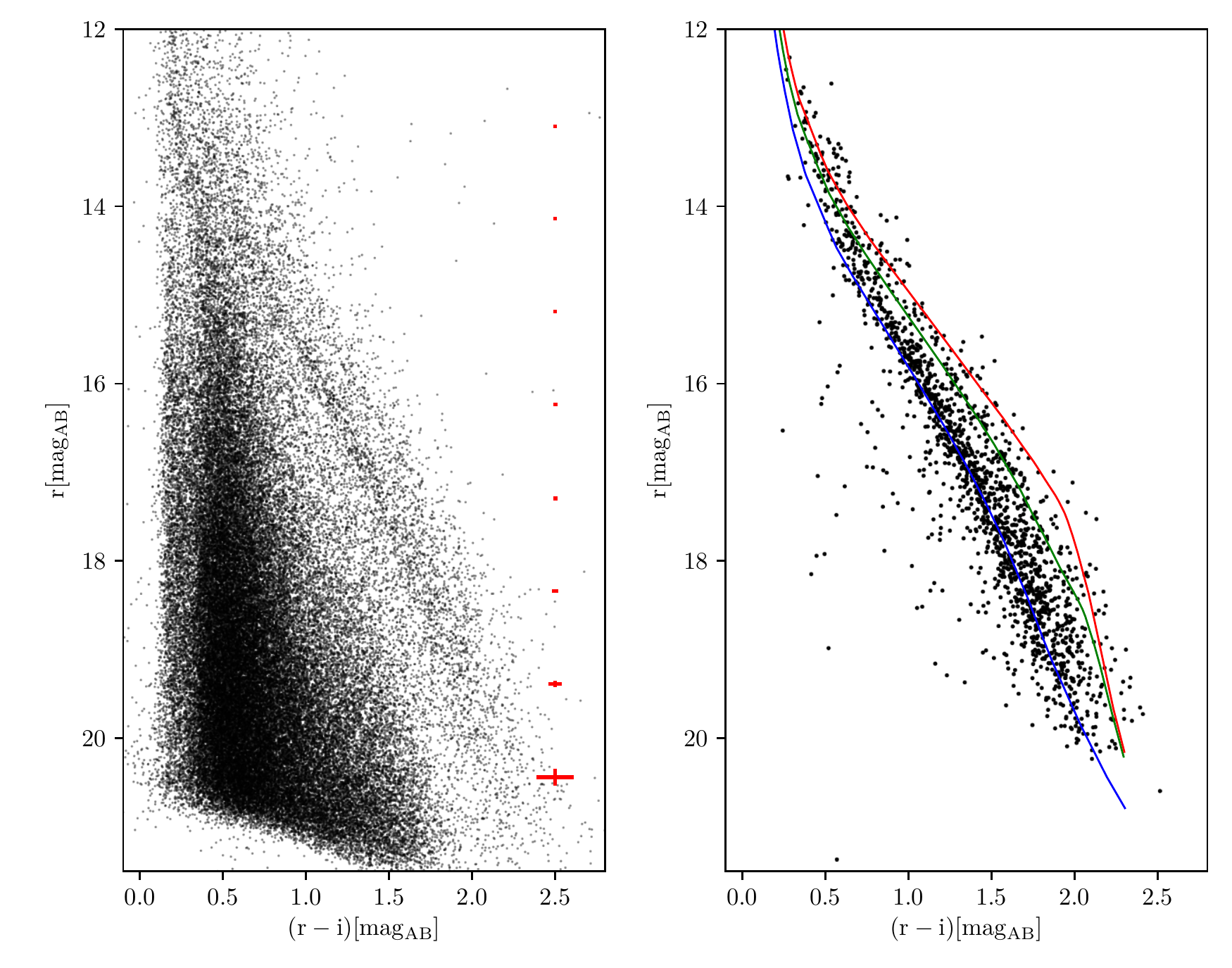} 
 \caption{Figure and its caption adapted from  \cite{Jerabkova2019}. Left panel: CMD for the initial catalog. The red crosses show 3−$\sigma$ color and magnitude errors. Right panel: The CMD of the parallax and proper motion selected ONC members. The color lines are the best fitting isochrones from the Pisa stellar evolutionary model of the three populations}
   \label{fig: fig1}
\end{center}
\end{figure*}

With these data in hand we were able to detect two well separated PMSs with a suggestion for the existence of a third one (see Fig.~\ref{fig: fig1}), despite facing incompleteness in the most centrally crowded region where the members of the second and third sequences are mainly located.

Our study supports the scenario that the PMS population of the ONC is composed of three  populations with different ages. 

\section{Feedback and stellar-dynamically regulated bursty star cluster formation }
\begin{figure*}[ht]
\begin{center}
 \includegraphics[width=5.2in]{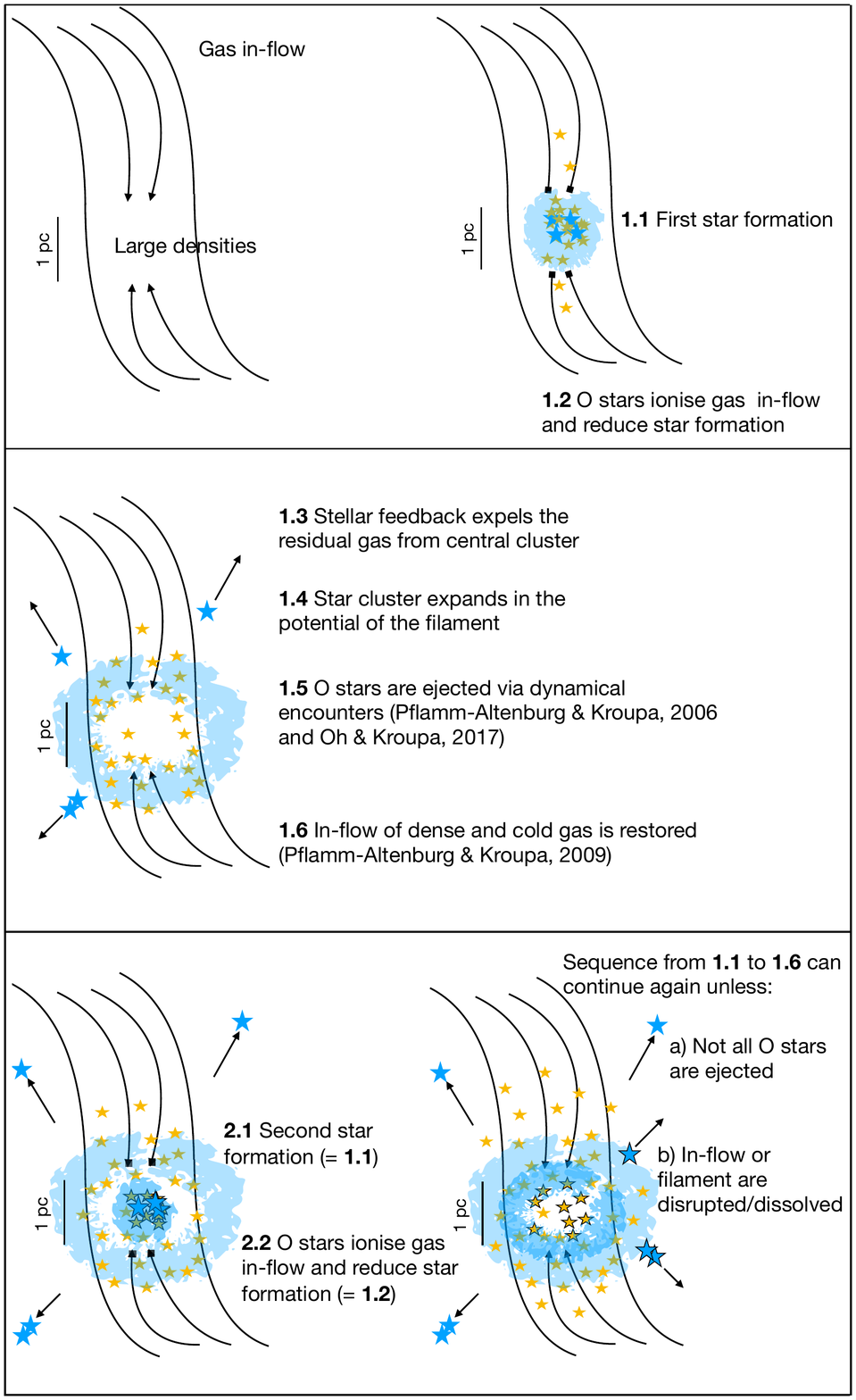} 
 \caption{Figure and its caption extracted from \cite{Kroupa2018}.  Schematics visualise the evolution of the molecular cloud filament, shown by the black curves, through the process of inflow (upper left panel), the monolithic formation of the first embedded cluster and termination of the inflow (upper right panel), the expansion of the first population due to gas expulsion and the ejection of the OB stars and resumption of gas inflow (middle panel), the monolithic formation of the second embedded cluster and termination of the inflow (lower left panel), the expansion of the second population due to gas expulsion and the ejection of its OB stars and resumption of gas inflow (lower right panel)}
   \label{fig: fig2}
\end{center}
\end{figure*}

\cite{Kroupa2018} presents a scenario for the formation of multiple co-eval populations separated in age by about 1 Myr as observed in the ONC. The scenario is explained in the schematic sketch, see Fig.~\ref{fig: fig2}. It is based on a converging inflow from a molecular 
gaseous filament that is building up a first stellar population. Once (and if) massive O stars 
are formed they ionise the inflow and suppress star formation in the cluster. However, the O stars 
can eject each on a short time scale ($<$ 1 Myr) -- before the converging filament is destroyed by their feedback. The inflow of molecular gas onto the cluster can then resume and a second stellar population can start forming. 

We show that for an ONC-like star cluster this process is realistic and thus can reproduce the observed three stellar populations. The mass-inflow history is constrained using this model and the number of OB stars ejected from each population is estimated for verification using Gaia data, also by explicit Nbody simulations by \cite{Wang2019}. As a further consequence of the proposed model, the three runaway O star systems, AE Aur, $\mu$ Col and $\eta$ Ori, are considered as significant observational evidence for stellar-dynamical ejections of massive stars from the oldest population in the ONC.

\section{Conclusions}
Using Gaia DR2 data we confirmed (\cite[Jerabkova et al. 2019]{Jerabkova2019})  the presence of several stellar populations in the ONC, as observed by \cite{Beccari2017}.

We suggested a theoretical model described in detail in \cite{Kroupa2018} that can account for these observations naturally by truncating star cluster formation via stellar feedback. However, stellar-dynamical ejections can likely remove the ionizing stars within about 1~Myr (\cite[Wang et al. 2019]{Wang2019}). The star cluster can then continue accretion of molecular material from the filament and form a new stellar population. This process can repeat until the ionizing stars are not ejected or the filament is exhausted.

Let us conclude with the remark that stellar dynamics might play an important role in the formation of star clusters, in addition to the filaments in molecular clouds.


\begin{thebibliography}{}

\bibitem[Beccari et al. (2017)]{Beccari2017}
{Beccari, G., Petr-Gotzens, M. G., Boffin, H. M. J., et al.} 2017,
\textit{A\&A}, 604, A22

\bibitem[Jeffries et al. 2011]{Jeffries2011}
{Jeffries, R. D., Littlefair, S. P., Naylor, T., \& Mayne, N. J. } 2011,
\textit{MNRAS}, 465, 2254


\bibitem[Jerabkova et al. (2019)]{Jerabkova2019}
{Jerabkova, T., Beccari, G., Boffin, H. M. J., Petr-gotzens, M. G., et al.} 2019,
\textit{A\&A}, 627, A57

\bibitem[Kroupa, Jerabkova et al. (2018)]{Kroupa2018}
{Kroupa, P., Jerabkova, T., Beccari, G., \& Yan, Z.} 2018,
\textit{A\&A}, 612, A74


\bibitem[Lada \& Lada 2003]{Lada2003}
{Lada, C. J. \& Lada, E. A.} 2003,
\textit{ARA\&A}, 41, 57


\bibitem[Oh \& Kroupa (2017)]{Oh2017}
{Oh, S., \& Kroupa, P.} 2017,
\textit{A\& A}, 590, A107

\bibitem[Palla \& Stahler 2000]{Palla2000}
{Palla, F. \& Stahler, S. W.} 2000,
\textit{ApJ}, 540, 255 


\bibitem[Pflamm-Altenburg \& Kroupa (2006)]{Pflamm2006}
{Pflamm-Altenburg, J., \& Kroupa, P.} 2006, 
\textit{MNRAS}, 373, 295


\bibitem[Wang et al. (2019)]{Wang2019}
{Wang, L., Kroupa, P., \& Jerabkova, T.} 2019,
\textit{MNRAS}, 484, 1843





\end{thebibliography}
\end{document}